\documentclass[runningheads]{svmult}

\usepackage{makeidx}   
\usepackage{graphicx}  
\usepackage{subeqnar}  
\usepackage{multicol}  
\usepackage{physprbb}  
\makeindex             

\def\arcsec{\hbox{$^{\prime\prime}$}}
\begin{document}
\title*{T-ReCS and Michelle: The Mid-Infrared Spectroscopic Capabilities of the Gemini Observatory}
\toctitle{T-ReCS and Michelle - The Mid-Infrared Spectroscopic \protect\newline Capabilities of the Gemini Observatory}
%
%
\titlerunning{Mid-IR Spectroscopic Capabilities of the Gemini Observatory}
%
\author{James M. De Buizer\inst{1}
\and R. Scott Fisher\inst{2}}
\authorrunning{De Buizer and Fisher}
%
%
\institute{Gemini Observatory, Southern Operations Center, Casilla 603
La Serena, Chile
\and Gemini Observatory, Northern Operations Center, 670 N. A'ohoku Place,
Hilo, Hawaii, 96720, USA}

\maketitle              

\begin{abstract}
Gemini Observatory's northern and southern telescopes are both presently being outfitted with facility mid-infrared imagers/spectrometers. This will allow observers the unique opportunity to apply to one observatory for all-sky spectroscopic access in the mid-infrared with the light gathering power of 8-meter telescopes. Gemini South has recently commissioned the Thermal-Region Camera and Spectrograph (T-ReCS) and is now available to perform queue observations for the community. T-ReCS is capable of low-resolution long-slit spectroscopy of R$\sim$100 near 10 and 20 $\mu$m, and medium-resolution long-slit spectroscopy of R$\sim$1000 near 10 $\mu$m. Gemini North is presently commissioning Michelle, which will be capable of low-, medium-, and high-resolution long-slit spectroscopy of R$\sim$200, 1000, and 3000, respectively, near 10 $\mu$m, as well as low-resolution long-slit spectroscopy of R$\sim$200 near 20 $\mu$m. Michelle can also perform echelle spectroscopy of R$\sim$30000 at 10 and 20 $\mu$m. The low-, medium-, and high-resolution spectroscopic modes of Michelle will be available to the public for the fall semester of 2004, and the echelle mode is expected to be available in 2005.
\end{abstract}

\section{Gemini's Twin Infrared-Optimized Telescopes}
The twin 8-m telescopes of Gemini Observatory are located in Hawaii in the northern hemisphere, and Chile in the southern hemisphere. Both telescopes are being outfitted with mid-infrared cameras/spectrometers, thus allowing observers the unique opportunity to apply for all-sky observations of mid-infrared targets. Another way in which Gemini sets itself apart from other 8-m class telescope facilities is its optimization for infrared observations.  

Ideally, a mirror coating should reflect as close to 100\% of the light that strikes it. In the thermal infrared region of the spectrum, however, a mirror coating emits a great deal of infrared radiation, and the statistical noisiness of this emission reduces an infrared instrument's sensitivity to astronomical sources. The emissivity of an optical surface is defined as the ratio between its level of emission and that of a perfect blackbody emitter, and it is roughly the inverse of the reflectivity. Therefore, in order to maximize infrared sensitivity, the mirror emissivity must be as low as possible. 

The Gemini telescopes employ single monolithic primary mirrors, rather than segmented ones whose intersegment gaps increase overall telescope emissivity. Furthermore, Gemini plans on coating primary and secondary mirror surfaces of both telescopes with silver (rather than the usual aluminum). Silver has a much lower emissivity than aluminum in the infrared, translating to unprecedented sensitivities for Gemini's infrared instruments. 

Gemini South's secondary mirror is already coated in silver, yielding a total system emissivity of 3.0\% at 9 $\mu$m. In May of 2004, the primary mirror of Gemini South will also be coated in silver, bringing the total emissivity down to an estimated 2.2\% or below. Gemini North also plans to be fully silver coated by the end of 2005. 

\section{T-ReCS}

\begin{table}[t]
\caption{Overview of T-ReCS Capabilities}
\begin{center}
\renewcommand{\arraystretch}{1.4}
\setlength\tabcolsep{15pt}
\begin{tabular}{@{}llp{1.8cm}l}
\hline\noalign{\smallskip}
\bf{Imaging:} & 20 filters in total \\
 & Diffraction-limited image quality: FWHM$\sim$0.25$\arcsec$ at 10 $\mu$m \\ 
 & Pixel size = 0.09$\arcsec$ (fixed)  \\
 & Field of view = 28.8$\arcsec$$\times$21.6$\arcsec$  \\
\noalign{\smallskip}
\hline
\noalign{\smallskip}
\bf{Spectroscopy:} & Slit widths = 0.26$\arcsec$ to 1.3$\arcsec$, Slit lengths = 21.6$\arcsec$   \\
 & Pixel size in spatial direction = 0.09$\arcsec$ (fixed) \\
 & \emph{Low-Resolution Spectroscopy:} \\
 & $\bullet$ R $\sim$ 100 near 10 and 20 $\mu$m  \\
 & $\bullet$ Dispersion at 10 $\mu$m = 0.019 $\mu$m/pix, $\Delta\lambda$ = 6 $\mu$m \\
 & $\bullet$ Dispersion at 20 $\mu$m = 0.029 $\mu$m/pix, $\Delta\lambda$ = 9 $\mu$m \\
 & \emph{Medium-Resolution Spectroscopy:} \\
 & $\bullet$ R $\sim$ 1000 near 10 $\mu$m  \\
 & $\bullet$ Dispersion at 10 $\mu$m = 0.0019 $\mu$m/pix, $\Delta\lambda$ = 0.6 $\mu$m  \\
\noalign{\smallskip}
\end{tabular}
\end{center}
\label{Tab1b}
\end{table} 

T-ReCS (Thermal-Region Camera and  Spectrograph) is a thermal infrared imager and long-slit spectrograph built by the University of Florida for Gemini South [3] . The imaging and spectroscopic modes of the instrument have already been commissioned and scientific queue data are already being collected. T-ReCS is proving to have high throughput and excellent image quality. 

The detector is a  Raytheon 320$\times$240 pixel Si:As IBC array with wavelength coverage of 7 to 26 $\mu$m. It has a switchable well capacity, allowing us to use a deep-well mode for imaging in wider filters or a medium-well mode for narrow-band imaging and spectroscopy. This is the largest available format mid-infrared array, and is the same type of detector used in Michelle. 

As summarized in Table 1, in its low-resolution spectroscopic mode, the T-ReCS array is large enough to encompass the full N-band or Q-band filter bandpasses. Because of the high dispersion of the medium-resolution grating, only small wavelength regions within the N band can be observed on the array at one time.   

\section{Michelle}

\begin{table}[t]
\caption{Overview of Michelle Capabilities}
\begin{center}
\renewcommand{\arraystretch}{1.4}
\setlength\tabcolsep{15pt}
\begin{tabular}{@{}llp{1.5cm}l}
\hline\noalign{\smallskip}
\bf{Imaging:} & 10 filters in total \\
 & Diffraction-limited image quality: FWHM$\sim$0.25$\arcsec$ at 10 $\mu$m \\ 
 & Pixel size = 0.10$\arcsec$ (fixed)  \\
 & Field of view = 32$\arcsec$$\times$24$\arcsec$   \\
\noalign{\smallskip}
\hline
\noalign{\smallskip}
\bf{Spectroscopy:} & Slit widths = 0.18$\arcsec$ to 1.44$\arcsec$, Slit length = 43.2$\arcsec$   \\
 & Pixel size in spatial direction = 0.18$\arcsec$ (fixed) \\
 & \emph{Low-Resolution Spectroscopy:} \\
 & $\bullet$ R $\sim$ 200 near 10 and 20 $\mu$m  \\
 & $\bullet$ Dispersion at 10 $\mu$m = 0.024 $\mu$m/pix, $\Delta\lambda$ = 7.7 $\mu$m \\
 & $\bullet$ Dispersion at 20 $\mu$m = 0.031 $\mu$m/pix, $\Delta\lambda$ = 10 $\mu$m  \\
 & \emph{Medium-Resolution Spectroscopy:} \\
 & $\bullet$ R $\sim$ 1000 near 10 $\mu$m  \\
 & $\bullet$ Dispersion at 10 $\mu$m = 0.0047 $\mu$m/pix, $\Delta\lambda$ = 1.5 $\mu$m  \\
 & \emph{High-Resolution Spectroscopy:} \\
 & $\bullet$ R $\sim$ 3000 near 10 $\mu$m  \\
 & $\bullet$ Dispersion at 10 $\mu$m = 0.0016 $\mu$m/pix, $\Delta\lambda$ = 0.5 $\mu$m  \\  
 & \emph{Echelle Spectroscopy:} \\
 & $\bullet$ R $\sim$ 30,000 near 10 $\mu$m  \\
 & $\bullet$ Dispersion varies - 1500 km/s at blaze wavelength  \\
\noalign{\smallskip}
\end{tabular}
\end{center}
\label{Tab1b}
\end{table}

Michelle is a mid-infrared echelle spectrograph and imager presently being shared between Gemini North [1] and UKIRT [2]. 

Like T-ReCS, the detector in Michelle is a  Raytheon 320$\times$240 pixel Si:As IBC array with wavelength coverage of 7 to 26 $\mu$m. It also has a switchable well capacity, and is used in deep-well mode for all imaging and medium-well mode for all spectroscopy.

Gemini North and is still undergoing characterization and commissioning of the spectroscopic modes of Michelle, however during the 2003B semester the instrument made its first imaging queue observations for the community. Michelle will be at UKIRT during 2004A, and afterwards it will return to Gemini for the foreseeable future.

The low-, medium-, and high-resolution spectroscopic modes of Michelle will be available to the public for semester 2004B, and the echelle mode is expected to be available in 2005.

\section{Conclusions}
With their versatile mid-infrared observing capabilities, T-ReCS and Michelle will empower astronomers with the tools needed to explore the nature of a broad range of astronomical objects and environments. The large infrared-optimized collecting area provided by the state-of-the-art Gemini telescopes, coupled with the large throughputs and excellent optics of T-ReCS and Michelle, are a powerful combination that will surely advance the field of mid-infrared astronomy.  

\section{Acknowledgements}
Gemini Observatory is operated by the Association of Universities for Research in Astronomy, Inc., under a cooperative agreement with the NSF on behalf of the Gemini partnership: the National Science Foundation (United States), the Particle Physics and Astronomy Research Council (United Kingdom), the National Research Council (Canada), CONICYT (Chile), the Australian Research Council (Australia), CNPq (Brazil) and CONICET (Argentina).


%

\end{document}